\documentclass{article}

\def\xxinput#1{\input#1}

\xxinput{vsolj02.sty}

\usepackage[dvipdfmx]{graphicx}
\usepackage[comma,colon]{natbib}

\usepackage[OT2,T1]{fontenc}

\def\cite{\citealt}
\setcitestyle{aysep={}}

\newcounter{author}
\def\altaffilmark#1{$^{#1}$}
\def\altaffiltext#1{$^{#1}$\,}

\def\authorcount#1#2{{\refstepcounter{author}\label{#1}
                     \altaffiltext{\ref{#1}}{#2}}}

\begin{document}

\begin{center}

\title{Long-lasting high state of the high-field polar AR UMa}

\author{
        Taichi~Kato\altaffilmark{\ref{affil:Kyoto}}
}

\authorcount{affil:Kyoto}{
     Department of Astronomy, Kyoto University, Sakyo-ku,
     Kyoto 606-8502, Japan \\
     \textit{tkato@kusastro.kyoto-u.ac.jp}
}

\end{center}

\begin{abstract}
\xxinput{abst.inc}
\end{abstract}

   There was an extremely soft X-ray
transient 1ES 1113$+$432 discovered from observations
by Imaging Proportional Counter of the Einstein Observatory
in 1979 \citep{rem94aruma}.  At that time, no optical object
brighter than $V$=16 within 2 arcmin of the X-ray location
was found.  The object was high above the Galactic plane
and a classical X-ray nova seemed unlikely.
\citet{rem94aruma} obtained optical imaging and found
a very blue object within 1.2 arcmin radius of the X-ray
error circle.  This optical counterpart ($V$=16.5)
had already been known as a semiregular (SR) variable
star \citep{GCVS,mei68aruma}, which was originally discovered
by \citet{hof63arruma} as an RR Lyr star (SON 7744) with
a range of 14.5--16 mag.
\citet{rem94aruma} obtained spectra, determined the orbital
period (1.9322~hr = 0.08051~d) and detected ``outburst''-like
variations on historical plates.
\citet{rem94aruma} identified this object to be a polar
(AM~Her star)
[see e.g., \citet{cro90polarreview} for a review of polars]
which spends most of the time in low-accretion states.
Referring to the manuscript of \citet{rem94aruma},
\citet{wen93aruma} re-examined the plates which were used
to give the earlier SR-type classification.  It turned out
difficult to determine the variability type in the 1960s
when the great diversity of cataclysmic variables (CVs)
was still unknown.  It is also worth noting that AR UMa
was not selected as a CV by ROSAT \citep{ROSATRXP},
which was/is known to be very efficient in discovering
polars.  AR UMa should have spent too much time in low states
to be detected even by ROSAT as a CV.  This object was then
shown to be the first high-field magnetic CV with
a field strengh of $\sim$230 MG \citep{sch96aruma}.

   While inspecting of ASAS-SN Sky Patrol Photometic Database
(ASAS-SN V2.0: \cite{ASASSNV2,ASASSN}), I noticed that AR UMa
had entered a long-lasting high state starting from 2022 October.
This brightening had been independently detected by
Yutaka Maeda on 2023 February 21 (14.1 mag, unfiltered CCD),
but apparently escaped detection by visual monitoring.
I immediately issued an alert (vsnet-alert 27706)\footnote{
   $<$http://ooruri.kusastro.kyoto-u.ac.jp/mailarchive/vsnet-alert/27706$>$.
}
on 2023 May 3, when ASAS-SN V2.0 observations had ended
on April 26.  After confirming that the Asteroid Terrestrial-impact
Last Alert System (ATLAS: \cite{ATLAS}) forced photometry
\citep{shi21ALTASforced} recorded this object still
bright in May, I issued a follow-up notice (vsnet-alert 27708)\footnote{
   $<$http://ooruri.kusastro.kyoto-u.ac.jp/mailarchive/vsnet-alert/27708$>$.
}.  Campaigns by the VSNET \citep{VSNET} and
VSOLJ teams (cf. vsnet-alert 27719)\footnote{
   $<$http://ooruri.kusastro.kyoto-u.ac.jp/mailarchive/vsnet-alert/27719$>$.
} and by the AAVSO\footnote{
   $<$https://www.aavso.org/aavso-alert-notice-824$>$.
} were launched.  Although time-resolved photometric data have
been reported, I primarily deal with the behavior of AR UMa leading
to the current bright state in this article.

\begin{figure*}
\begin{center}
\includegraphics[width=16cm]{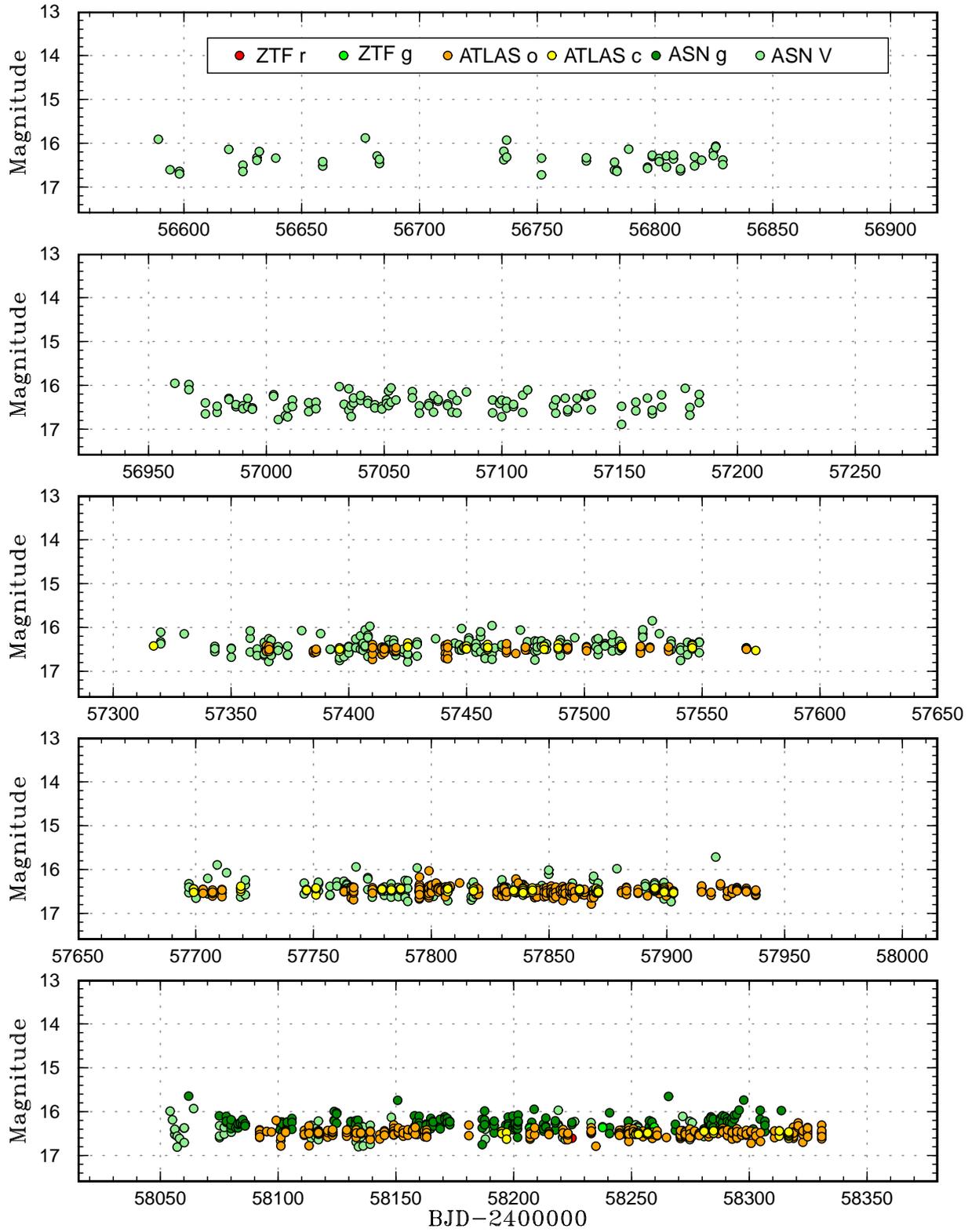}
\caption{
   Light curve of AR UMa in 2013--2018.
The object was in low state.  The symbols used in
figure \ref{fig:lc2} are also displayed.
}
\label{fig:lc1}
\end{center}
\end{figure*}

\begin{figure*}
\begin{center}
\includegraphics[width=16cm]{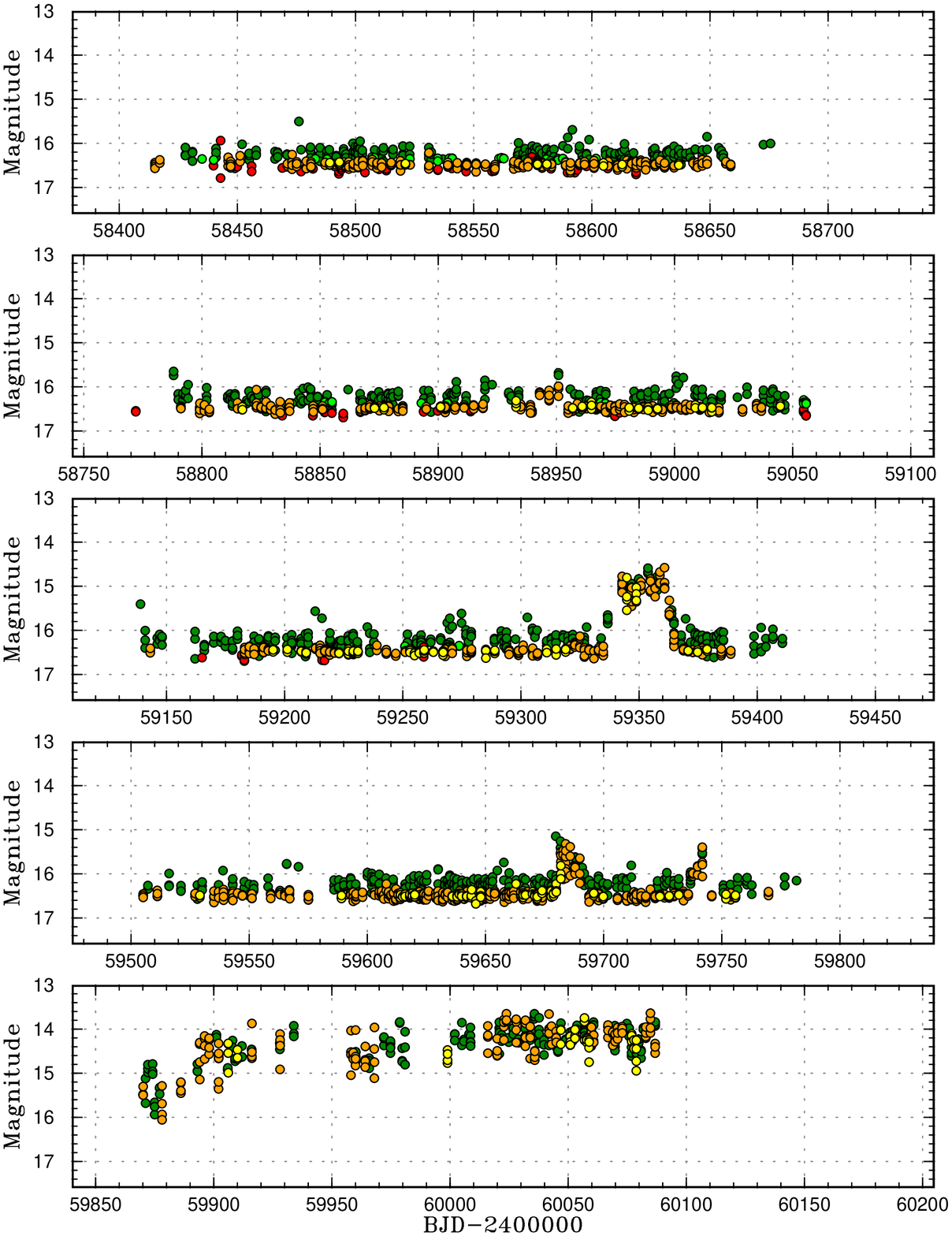}
\caption{
   Light curve of AR UMa in 2018--2023.
The symbols are the same as in figure \ref{fig:lc1}.
Transient high states were seen in the third and fourth
panels.  The object entered a long-lasting high state
in the last panel.
}
\label{fig:lc2}
\end{center}
\end{figure*}

   The light curve since 2013 using ATLAS, ASAS-SN (not V2.0) and
Zwicky Transient Facility
(ZTF: \cite{ZTF})\footnote{
   The ZTF data can be obtained from IRSA
$<$https://irsa.ipac.caltech.edu/Missions/ztf.html$>$
using the interface
$<$https://irsa.ipac.caltech.edu/docs/program\_interface/ztf\_api.html$>$
or using a wrapper of the above IRSA API
$<$https://github.com/MickaelRigault/ztfquery$>$.
} observations is shown in figures
\ref{fig:lc1} and \ref{fig:lc2}.
Several ASAS-SN observations were omitted as noises
by comparing with observations on the same night.
Before this, there had been some observations reported to
VSOLJ and VSNET.  There was no evidence for a long-lasting
high state in 2009--2013, although observations were
fragmentary.  There were short-lived high states in
2007 January to March and in 2008 January, both detected
by Ian Miller by snapshot CCD observations
(see also \cite{sch99aruma} for the light curve).
Observations before them were not ideal and visual
observations since 1997 reporting magnitudes of 15.0--16.0
probably referred to the low state.  There was a documented
short-lived high state in 1996 December \citep{sch99aruma}.
It appears that the high state starting from 2022 October
is the first long one since the discovery.
According to \citet{wen93aruma}, low states
were predominant and only six high states since 1961
(spanning 32 years) had been detected almost like
dwarf nova outbursts.  There may have not been
a long-lasting high state even since the discovery of
this object as a variable star.

   There were three short-lived brightening events in 2021 May,
2022 April and likely in 2022 June (figure \ref{fig:lc2})
despite the long absence (nearly 8~yr) of a similar event
before them.  These short-lived events may have been precursors
to the present high state.  Although it has been argued that
long-lasting low states already challenge the theories
\citep{bre12j1250j1514}, whether these short-lived brightening
events reflected gradually increasing mass-transfer
or such precursors could initiate some sort of positive
feedback to maintain the high state may be an interesting
theoretical topic.

\begin{figure*}
\begin{center}
\includegraphics[width=14cm]{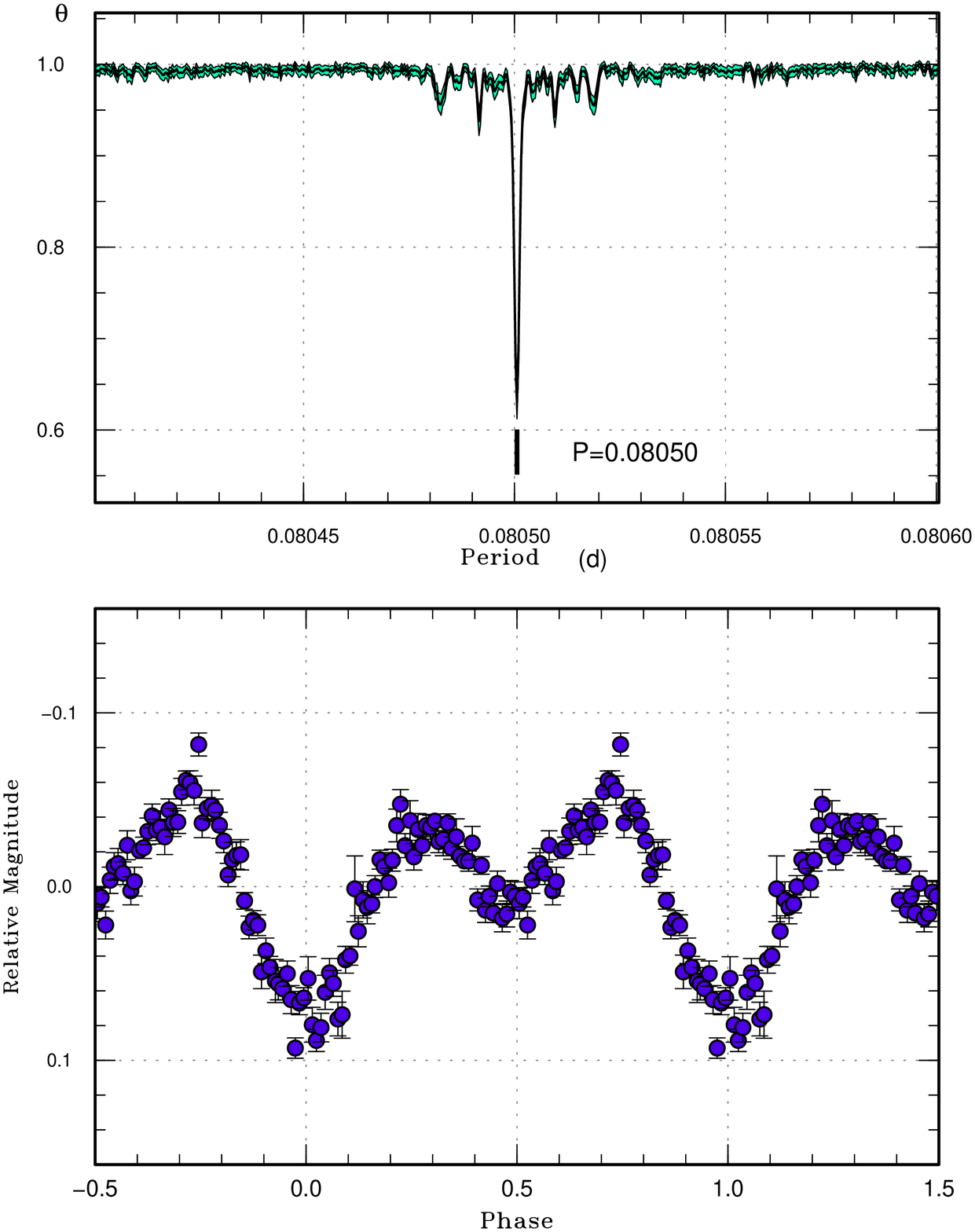}
\caption{
   Orbital profile of AR UMa in low states.
   (Upper): PDM analysis.  The bootstrap result using
   randomly contain 50\% of observations is shown as
   a form of 90\% confidence intervals in the resultant 
   $\theta$ statistics.
   (Lower): Phase plot.
}
\label{fig:lowpdm}
\end{center}
\end{figure*}

   The orbital profile during the low states before 2022
October is shown in figure \ref{fig:lowpdm}.
The ASAS-SN, ATLAS and ZTF data were combined
after removal of the long-term trends by
locally-weighted polynomial regression (LOWESS: \cite{LOWESS}).
The period wes determined using the phase dispersion
minimization (PDM: \cite{PDM}) method, whose error was
estimated by the methods of \citet{fer89error,Pdot2}.
Strong ellipsoidal variations are apparent
[see \citet{sch96aruma,how01aruma}].
The difference between the primary and secondary
maxima was much smaller in \citet{sch96aruma} and \citet{how01aruma}
because they used infrared bands.
In drawing the phase-averaged light curve, I used
the following ephemeris determined from these data:
\begin{equation}
T_0\mathrm{(BJD)} = 2458786.470(1) + 0.08050059(2) E.
\label{equ:phase}
\end{equation}
This period agrees with the one 0.08050074(12)~d
by \citet{sch99aruma} from radial-velocity observations.
The difference in the zero phases between this
and \citet{sch99aruma} (zero crossing by radial-velocity
observations) corresponds to a 0.076 orbit after 103304
orbital cycles.  Assuming that the present photometric
minimum is equivalent to the zero crossing phase,
the improved orbital period is 0.08050066(1)~d, which
can be considered as the current best determination of
the period of this binary.

\begin{figure*}
\begin{center}
\includegraphics[width=14cm]{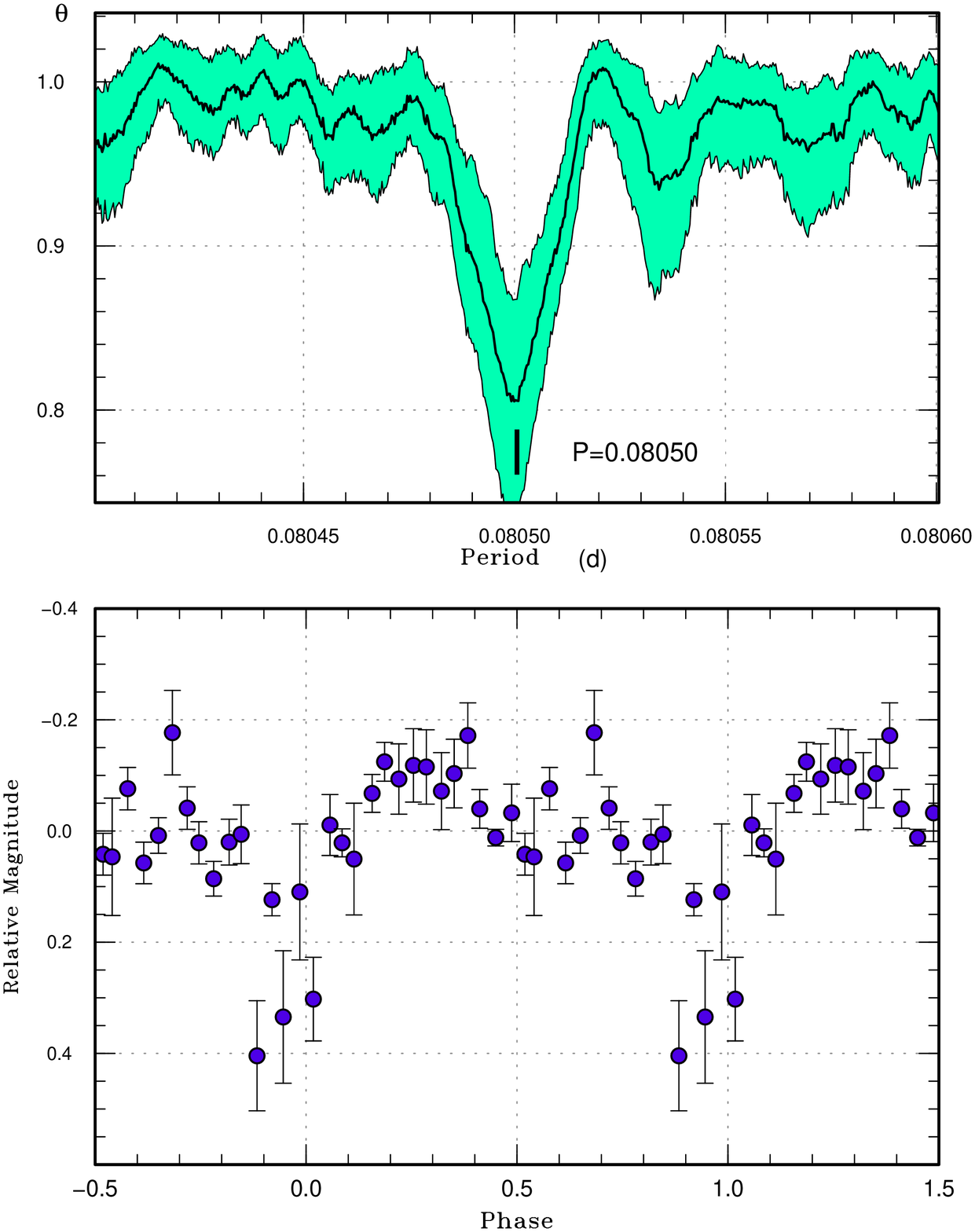}
\caption{
   Orbital profile of AR UMa in the current high state.
   (Upper): PDM analysis.  (Lower): Phase plot.
}
\label{fig:highpdm}
\end{center}
\end{figure*}

   The orbital profile during the current high state
is shonw in figure \ref{fig:highpdm}.  ZTF data for this
interval are not yet publicly available and ATLAS and
ASAS-SN data were combined.  Although the statistic is
worse than the low-state profile, a bump at phase 0.2--0.3
appeared.

\section*{Acknowledgements}

This work was supported by JSPS KAKENHI Grant Number 21K03616.
I am grateful to Naoto Kojiguchi for helping downloading
ASAS-SN Sky Patrol Photometic Database and ZTF data and to
ASAS-SN, ATLAS and ZTF teams for making their data available
to the public.
I am grateful to Franz-Josef Hambsch, Tam\'as Tordai,
Tonny Vanmunster, Filipp D. Romanov and Hiroshi Itoh
for providing time-resolved CCD photometry.  Although they
were not presented in this publication, they were useful in
confirming the profile and phasing of the high-state variation.

This work has made use of data from the Asteroid Terrestrial-impact
Last Alert System (ATLAS) project.
The ATLAS project is primarily funded to search for
near earth asteroids through NASA grants NN12AR55G, 80NSSC18K0284,
and 80NSSC18K1575; byproducts of the NEO search include images and
catalogs from the survey area. This work was partially funded by
Kepler/K2 grant J1944/80NSSC19K0112 and HST GO-15889, and STFC
grants ST/T000198/1 and ST/S006109/1. The ATLAS science products
have been made possible through the contributions of the University
of Hawaii Institute for Astronomy, the Queen's University Belfast, 
the Space Telescope Science Institute, the South African Astronomical
Observatory, and The Millennium Institute of Astrophysics (MAS), Chile.

Based on observations obtained with the Samuel Oschin 48-inch
Telescope at the Palomar Observatory as part of
the Zwicky Transient Facility project. ZTF is supported by
the National Science Foundation under Grant No. AST-1440341
and a collaboration including Caltech, IPAC, 
the Weizmann Institute for Science, the Oskar Klein Center
at Stockholm University, the University of Maryland,
the University of Washington, Deutsches Elektronen-Synchrotron
and Humboldt University, Los Alamos National Laboratories, 
the TANGO Consortium of Taiwan, the University of 
Wisconsin at Milwaukee, and Lawrence Berkeley National Laboratories.
Operations are conducted by COO, IPAC, and UW.

The ztfquery code was funded by the European Research Council
(ERC) under the European Union's Horizon 2020 research and 
innovation programme (grant agreement n$^{\circ}$759194
-- USNAC, PI: Rigault).

\section*{List of objects in this paper}
\xxinput{objlist.inc}

\section*{References}

We provide two forms of the references section (for ADS
and as published) so that the references can be easily
incorporated into ADS.

\newcommand{\noop}[1]{}\newcommand{\hyphalt}{-}

\renewcommand\refname{\textbf{References (for ADS)}}

\xxinput{arumaaph.bbl}

\renewcommand\refname{\textbf{References (as published)}}

\xxinput{aruma.bbl.vsolj}

\end{document}